\documentclass[a4paper,11pt]{article}
 \usepackage{fullpage}
\usepackage{array}
\usepackage[latin1]{inputenc}
\usepackage[english]{babel}
\usepackage{graphicx}
\usepackage{placeins}
\usepackage{amssymb,amsmath}
\usepackage{color,listings}
\usepackage{subfig}

\definecolor{light-gray}{gray}{0.9}
\title{Deployment and Evaluation of a 802.15.4 Heterogeneous Network} 

\author{Orestis Akribopoulos, Vasileios Georgitzikis, Christos Koninis, Ioannis
Papavasileiou\\ and Ioannis Chatzigiannakis\\
Research Academic Computer Technology Institute (RACTI) and\\
Computer Engineering and Informatics Department, University of Patras, Greece\\
Email:\{akribopo, tzikis, koninis, papabasile, ichatz\}@cti.gr}

\begin{document}

\maketitle

\section{Introduction}
\label{sec:intro}
In this work we study the performance of a heterogeneous wireless sensor network
which consists of 4 different hardware platforms
(TelosB, SunSPOT, Arduino, iSense). These hardware platforms  are the most
representative ones, as used by the relevant research community. 

All hardware platforms use 802.15.4 compliant radios. Due to partial implementation of the standard,
they do not communicate out of the box. A first contribution of our work is a careful description of the necessary steps to make such a heterogeneous network
interoperate. Our software code is available online. 

We deploy a heterogeneous network testbed and conduct a thorough evaluation of the performance.
We examine various network performance metrics (e.g., transmission rate, receiving rate, packet loss, etc.), and assess the
capabilities of each device and their intercommunication. We used different
setups (e.g., distance between transmitters and receivers, etc.) to better understand 
the network limitations for each hardware platform.

Out study demonstrates the differences between the hardware platforms. The
platform with the larger transmitting and receiving rate is the iSense. iSense
is the only device  with 0\% packet loss regardless of the transmitting device,
the distance and the packet size. In contrast to iSense, Arduino is the only
device with packet loss over 50\% in large distances, with packet size over 50
bytes. The SunSPOT and TelosB use the same hardware. Interestingly, although the
SunSPOT has a much more powerfull processor and a lot of more memory
available, the way Java Virtual Machine operates, limits the overall network
performance. Thus the TelosB, achieves much higher transmission rates. We also observe that the RSSI provided by each hardware platform are totally different
and in fact no correlation can be found. This is an alarming observation for Network Algorithms that rely on RSSI and need to operate in heterogeneous setting.

\section{Hardware Description}

\subsection{Arduino Duemilanove with XBee 1mW Chip Antenna}
Arduino~\cite{arduino}(figure~\ref{arduino}) is an open-source electronics
prototyping platform. The Arduino
Duemilanove is a microcontroller board based on the ATmega328. It has 14 digital
input/output pins, 6 analog inputs, a 16 MHz crystal oscillator, a USB
connection, a power jack, an ICSP header, and a reset button. The Arduino is
connected with the XBee~\cite{xbee} Series 1 Chip Antenna. The Xbee module
provides IEEE
802.15.4 network connectivity to the Arduino.
The core libraries of Arduino are written in C and C++ and compiled using avr-gcc and AVR Libc.
Arduino hardware is programmed using a Wiring~\cite{wiring} based language, similar to C++ with some simplifications and modifications, 
and a Processing~\cite{processing} based IDE.
Libelium Waspmote~\cite{waspmote} has a similar microcontroller and a similar
radio transmitter to Arduino.

\subsection{SunSPOT}
SunSPOT~\cite{sunspot}(figure~\ref{sunSpot}) is a small,
battery-operated device running the Squawk Java Virtual Machine, which acts as
both an operating system and a software application platform, 
allowing programming of the devices in the Java Micro Edition (J2ME) platform.
It uses an $180$MHz ARM 9 processor with $512$KB of RAM and $4$MB Flash. An IEEE
802.15.4 compliant CC2420 Chipcon transceiver is used for communication. 

\subsection{TelosB mote}
\begin{figure}[!ht]
  \centering
\subfloat[Arduino]{\label{arduino}\includegraphics[width=0.18\textwidth]{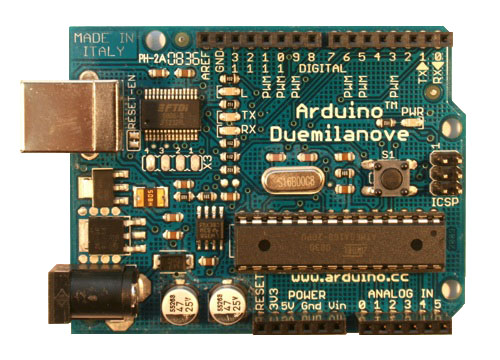}}                
\subfloat[SunSPOT]{\label{sunSpot}\includegraphics[width=0.15\textwidth]{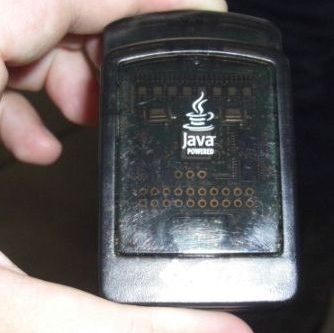}}
\subfloat[TelosB]{\label{telosb}\includegraphics[width=0.18\textwidth]{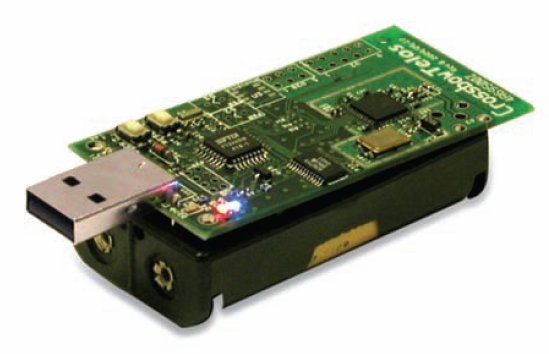}}
\subfloat[iSense]{\label{core-module}\includegraphics[width=0.11\textwidth]{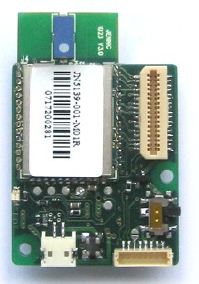}}
  \caption{Hardware Devices}
\end{figure}

TelosB~\cite{telosb}(figure~\ref{telosb}) is 2.4GHz IEEE 802.15.4 compliant and
comes with a 10KB RAM module, a 48 kBytes program Flash memory and
the TI-MSP430 microprocessor, which is running on 4 MHz. TelosB are running the TinyOS
version 2.1.0~\cite{tinyos} operating system and their software is written 
in nesC. Prisma Sesnse Quax MS-Pro~\cite{prismasense} is a
hardware platform with similar processor to TelosB and the ScatterWeb
MSB-430~\cite{scatter} is another platform with similar architecture and the
same processor but uses a different radio chipset (CC1020) which is not 802.15.4
compliant.

\subsection{iSense}
iSense~\cite{isense} was obtained by Coalesenses GmbH
based in Luebeck, Germany. It is comprised by iSense Core modules
(figure~\ref{core-module}) for both Receiving and Transmitting Nodes.
The Core module uses IEEE 802.15.4 Zigbee compliant radio(JN5139) and has a
32bit RISC Controller running at 16MHz. It features 96kB of RAM abd 128kB of
Serial Flash.
The iSense OS and the programming of the devices is in the  C++ language.

The hardware differences are summarized in the following table, Table~\ref{diffs}.

\begin{center}
\begin{table}[ht!]
\begin{tabular}{l|crrrcc}
 & \textbf{Processor} & \textbf{MIPS} & \textbf{RAM} & \textbf{Flash} & \textbf{Radio} & \textbf{Program. Lang.}\\
\hline
\hline
\textbf{Arduino} & ATmega328(16 MHz) & 16 & 16 KB  & 32 KB  & XBee Series 1 & Wiring (C++)\\
\textbf{SunSPOT} & ARM920T(180 MHz) & 200 & 512 KB & 4 MB & CC2420 & J2ME\\
\textbf{TelosB} & MSP430(16 MHz) & 16 & 10 KB & 48 KB & CC2420 & nesC\\
\textbf{iSense} & JN5139(16 MHz) & 16 & 96 KB  & 128 KB & JN5139& C++

\end{tabular}
\caption{\label{diffs} Comparison of Platforms.}
\end{table}
\end{center}

\section{MAC Implementation}
Each of the considered sensor devices is equipped with an IEEE std. 802.15.4 -
2003~\cite{802}
compliant radio to perform wireless communication.
However, each one of them provides only partial implementations of the IEEE
802.15.4, which are not compatible with each other. Therefore the
communication between the 4 different devices is not possible out of the box.

IEEE 802.15.4 provides two different addressing modes, the 16-bit
addressing and the 64-bit.
The SunSPOT radio stack supports only the 64-bit addressing mode, while TelosB
supports only the 16-bit.
Radio stacks of XBee and iSense provide both the 64-bit and the 16-bit
addressing modes.

The first step on our Heterogeneous Sensor Network was to set all the devices to
the 16-bit addressing mode. XBee was set on the 802.15.4 Mac mode with
auto-ACKs.
We also implemented a new radio stack on SunSPOT which supports the 16-bit
addressing mode. 

Based on the LowPAN specification, the Sun SPOT library provides routing,
meshing and fragmentation using the LowPAN, on the network layer. LowPAN adds
some extra headers on the 802.15.4 packets. In particular, after the 802.15.4
headers, two extra bytes are added by the LowPAN which define whether the packet
is LowPAN compliant, whether it is fragmented, whether it is meshed etc. 

Our network stack does not support fragmentation and mesh routing. So on each
radio stack, two constant bytes at the beginning of the payload of each packet
had to be added, in order to define that each packet is not fragmented or meshed. 
In this way, the LowPAN on the network layer of the SunSPOT is bypassed and the
communication between the 4 different sensor nodes is possible.

The differences of the 4 platforms are summarized in the following table, table~\ref{sum}  

{%
\newcommand{\mc}[3]{\multicolumn{#1}{#2}{#3}}
\begin{center}
\begin{table}[ht!]
\begin{center}
\begin{tabular}{ccccc}
\textbf{Platform}  &  \textbf{Max Payload Size} & \mc{2}{l}{\textbf{Addressing Mode}} &  \textbf{Incompatibilities}\\
 &  &  \textbf{16-bit} & \textbf{64-bit} & \\
\hline 
\hline 
Arduino XBee & 100 bytes & YES & YES & Extra Headers (MaxStream Headers)\\
SunSPOT & 113 bytes & NO & YES & Extra Headers (LowPan)\\
TelosB & 128 bytes & YES & NO &  Auto Ack is Disabled\\
iSense & 116 bytes & YES & YES & \\
\hline 
\end{tabular}

\caption{\label{sum} Comparison of Platforms.}
\end{center}
\end{table}
\end{center}
}%

The customized radio stack for SunSPOT was implemented in Java J2ME, while the library, which enables the communication on iSense and Arduino, was implemented in C++.
TelosB motes were running the TinyOS version 2.1.0~\cite{tinyos}, so the component for the Telosb was written in nesC.

A similar work, is the TinySPOTComm~\cite{tinyspot} library. In contrast to our work, TinySPOTComm enables the communication between only 2 devices, SunSPOT and TelosB.

\section{Experimental Setup}
The setup consists of two different group of sensor nodes: the Receiving
Nodes and Transmitting Nodes. Each of these groups consists of 4 nodes, one
Arduino, one
SunSPOT, one TelosB and one iSense. The layout of the experiment  can be seen in
the following figure~\ref{layout}.
\begin{figure}[!ht]
\centering
\includegraphics[width=0.95\textwidth]{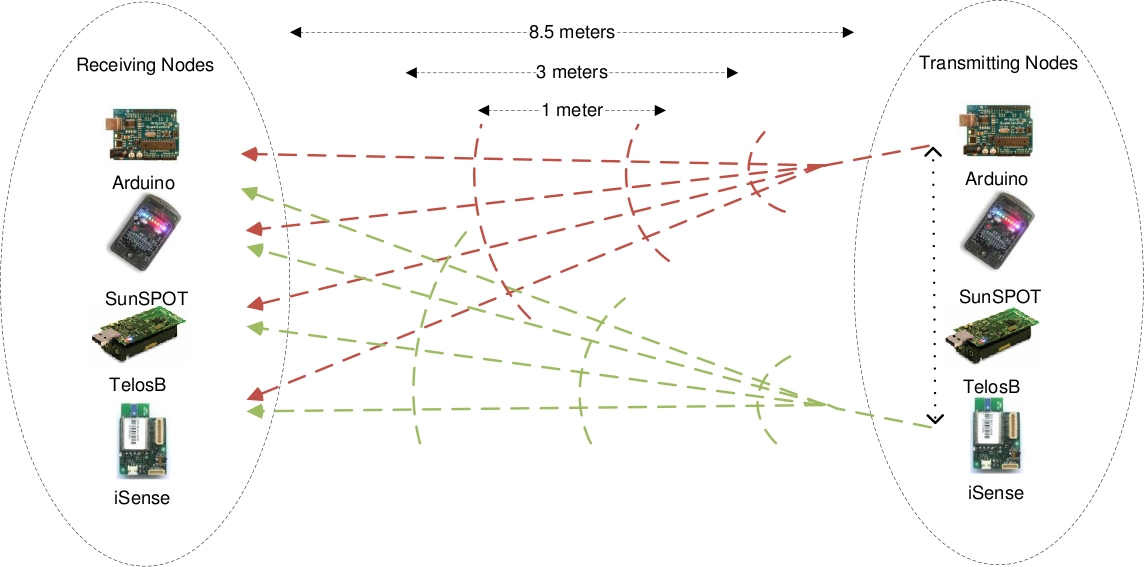}
\caption{\label{layout} The layout of the experiment.}
\end{figure}

All nodes were positioned at a height of 60cm from the ground to avoid
ground reflections of the wireless signal. All four Receiving Nodes
had the same orientation, pointing their antennas to the Transmitting Nodes (in contrast to figure~\ref{layout}). 
We use 3 different setups regarding the distance between the Transmitting nodes
and the Receiving Nodes. The Transmitting Nodes were placed in 3 different
distances from the Receiving Nodes: 1 meter, 3 meters and 8.5 meters.

For all experiments we measure the following performance metrics:
\begin{description}
 \item \textit{Received Packets per Second:}  is the number of packets received in
one second from a Receiving Node. This metric can be used (in combination to the payload size) to compute the effective throughput.
 \item \textit{Packet Loss}: defines the percentage of packets which fail to reach the
Receiving Node.
 \item \textit{Received Signal Strength Indication (RSSI):} RSSI is the relative received
signal strength in a wireless environment, in arbitrary units. RSSI is an
indication of the power level being received by the antenna. Therefore, the
higher the RSSI number (or less negative in some devices), the stronger the
signal.
\end{description}

On each repetition, one of the 4 Transmitting Nodes was broadcasting 500 beacon
messages
with payload size varying from 6 to 96 bytes. Each Receiving Node was recording the RSSI and the received timestamp of each
beacon as well as the total number of the received packets. Since the beacon was a broadcast
packet, the 4 Receiving Nodes were simultaneously receiving these beacons. Each one of the 4 Transmitting Nodes was placed at 3 different distances
from the Receiving Nodes (i.e., 1, 3, 8.5 meters). After the
measurements have been conducted the average values for 20 payload sizes for each distance have
been calculated. 

\section{Experimental Results}
Here we present the results of our experiment for the 3 different distances
between the Transmitting and the Receiving Stations.

\subsection{Broadcasting Rate}
\begin{figure}[!ht]
\centering
\includegraphics[width=0.6\textwidth]{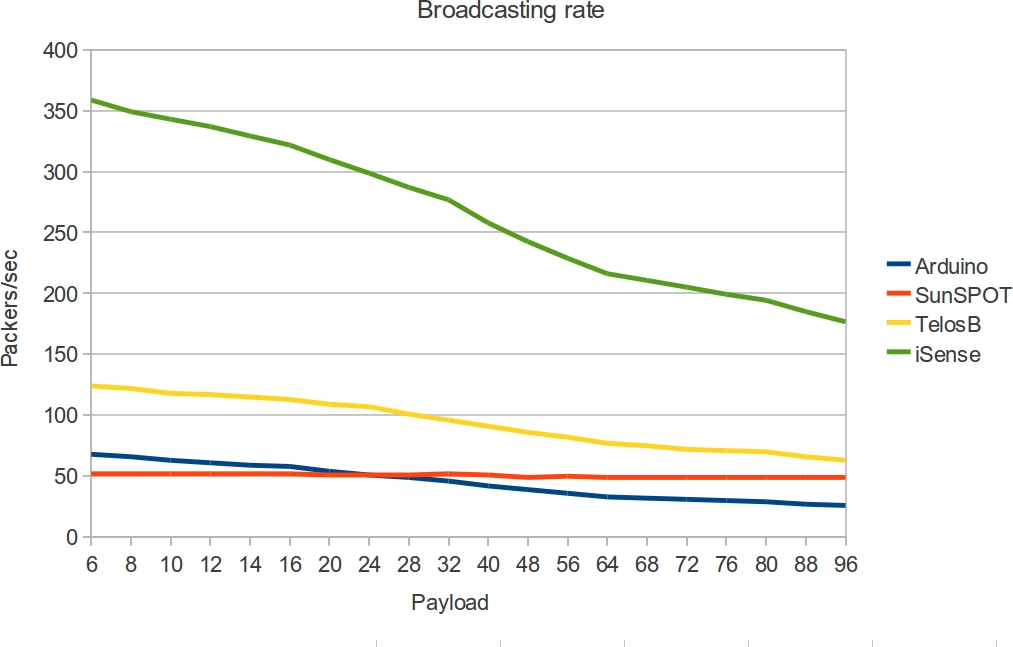}
\caption{\label{ppsOUT} Packets per Second on Broadcasting Nodes}
\end{figure}

We start by measuring the broadcasting rate of the 4 different devices, for
packets of different message sizes (i.e., message payload).
We do this by continuously sending, from each of the Transmitting Nodes, 20
different payload sizes, ranging from 6 to 96 bytes. 
We calculate, for each device, the total period of time required to transmit 500
packets. We repeat the experiment 9 times to achieve good average results.
The average results are illustrated in the figure~\ref{ppsOUT}.

It should be mentioned that the XBee modules are connected with the Arduino
using a serial UART interface which supports up to 115200 baud rate. But above
the threshold of 38400 the communication between the Arduino and the XBee module
becomes unstable in high transmission rates. In particular, when the received packets per second on XBee are above 150, Arduino continuously restarts. In our experiment the baud rate was
set on 38400.

The device with the larger broadcasting rate is the iSense, while the SunSPOT and Arduino have the smaller broadcasting rate.
The broadcasting rate on TelosB is almost double the rate on SunSPOT, because the Squawk Java Virtual Machine limits the performance of the SunSPOT.

\subsection{Received Packets per Second}
Each one of the 4 Transmitting Nodes was broadcasting 500 beacon messages
with payload size varying from 6 to 96 bytes. The received packets per second depend on the transmission rate of Transmitting Node. When  the Transmitting Node is a SunsPOT or an Arduino,
the received pps is almost the same for all  Receiving Nodes and is equal with the transmission rate of SunSPOT and Arduino respectively. 
When TelosB is the Transmitting Node, the received pps is the same for all Receiving nodes, except from Arduino. The received pps on Arduino is decreasing for payload sizes over 28 bytes.

\begin{figure}[!ht]
\centering
\includegraphics[width=0.32\textwidth]{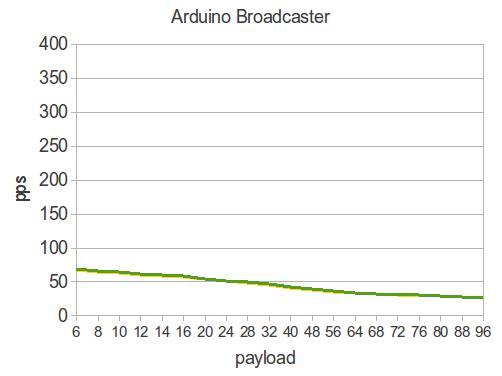}
\includegraphics[width=0.32\textwidth]{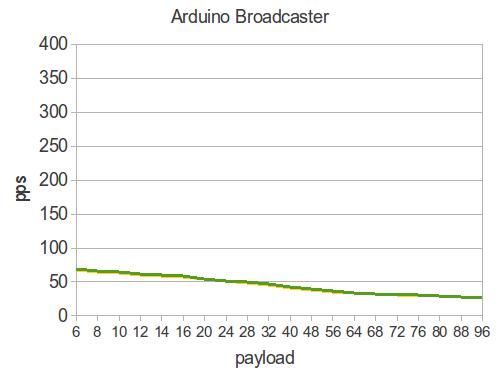}
\includegraphics[width=0.32\textwidth]{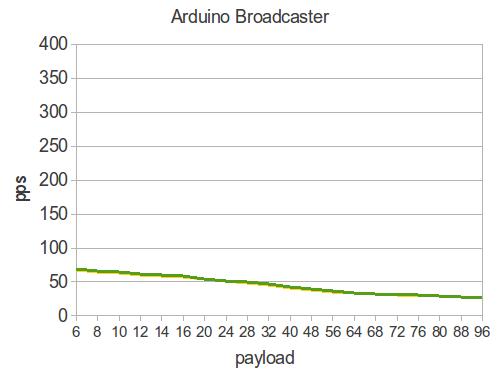}
\includegraphics[width=0.32\textwidth]{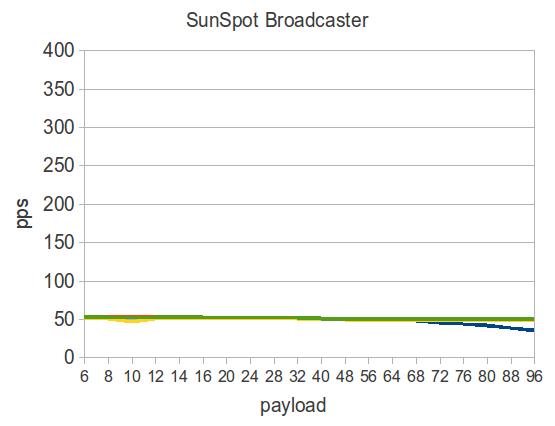}
\includegraphics[width=0.32\textwidth]{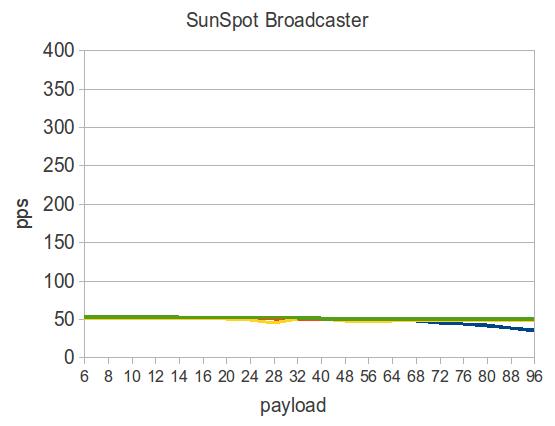}
\includegraphics[width=0.32\textwidth]{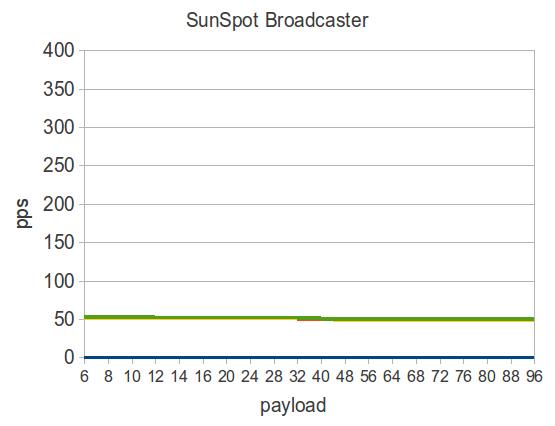}
\includegraphics[width=0.32\textwidth]{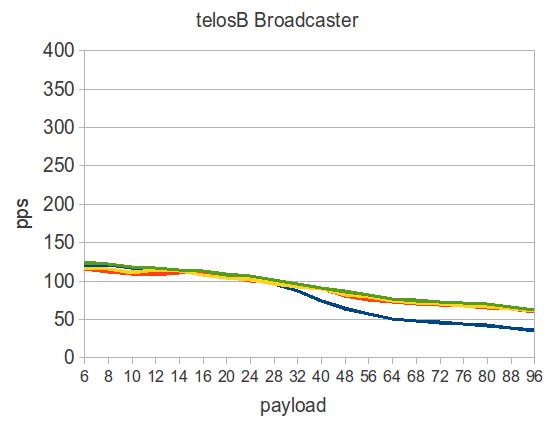}
\includegraphics[width=0.32\textwidth]{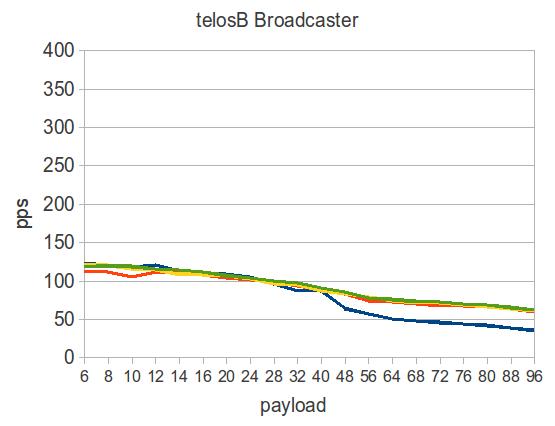}
\includegraphics[width=0.32\textwidth]{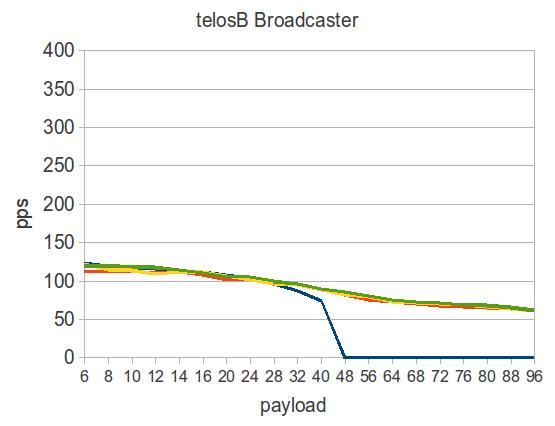}
\includegraphics[width=0.32\textwidth]{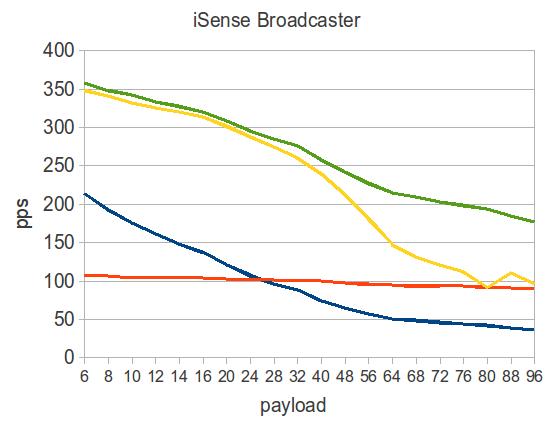}
\includegraphics[width=0.32\textwidth]{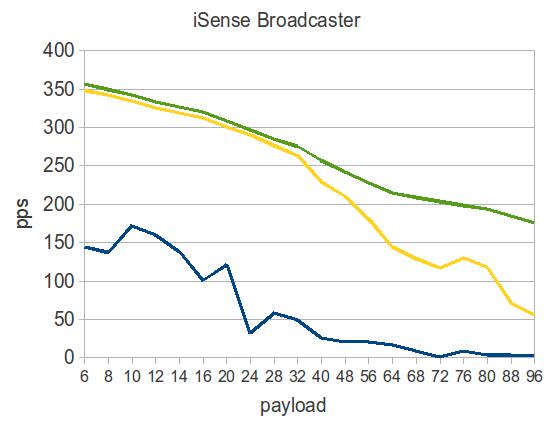}
\includegraphics[width=0.32\textwidth]{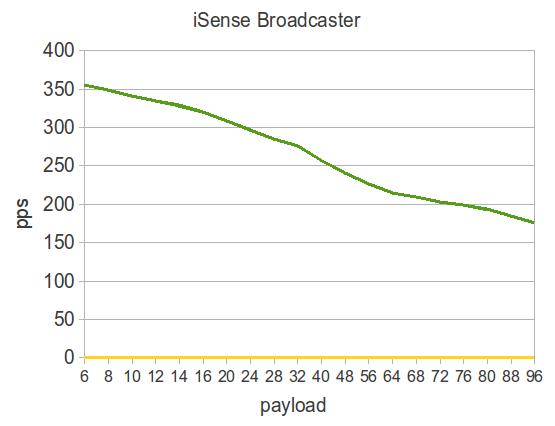}

\begin{tabular}{ m{0.1\textwidth} m{0.30\textwidth}  m{0.30\textwidth} m{0.32\textwidth}}
  & \textbf{1 meter} &\textbf{3 meters} & \textbf{8.5 meters}
\end{tabular}
\includegraphics[width=0.4\textwidth]{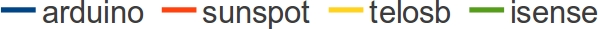}
\caption{\label{pps} Packets per Second on Receiver}
\end{figure}

In the following figure, the Received Packets per Second rates are illustrated, figure~\ref{pps}.
The distance between the Transmitting  and the Receiving Nodes on the left column, on the middle column and in the right column, is 1 meter, 3 meters and 8.5 meters respectively.

\newpage

\subsection{Packet loss rate}
The packet loss rate is larger while the distance is increasing. We also observe an increased packet loss rate, when the Transmitting Node has a broadcasting rate larger than the maximum receiving rate on the Receiving Node. 

\begin{figure}[!ht]
\centering
\includegraphics[width=0.32\textwidth]{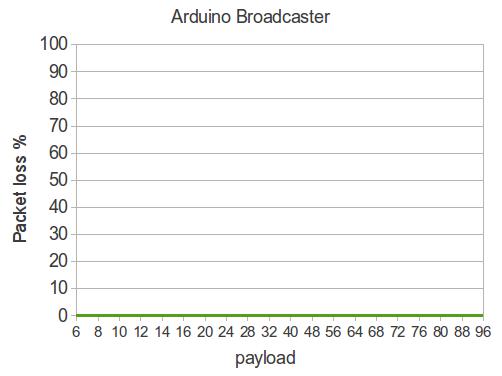}
\includegraphics[width=0.32\textwidth]{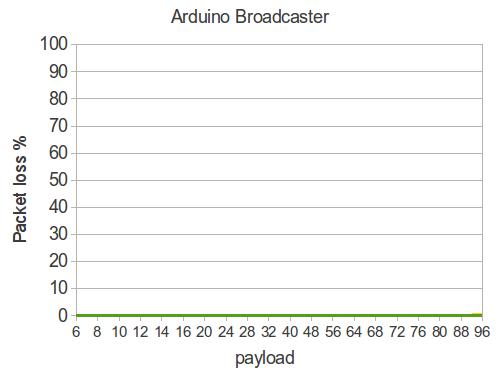}
\includegraphics[width=0.32\textwidth]{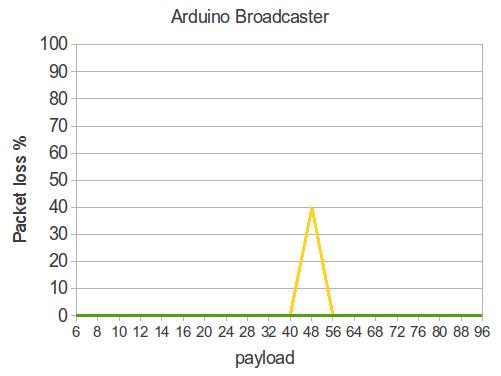}
\includegraphics[width=0.32\textwidth]{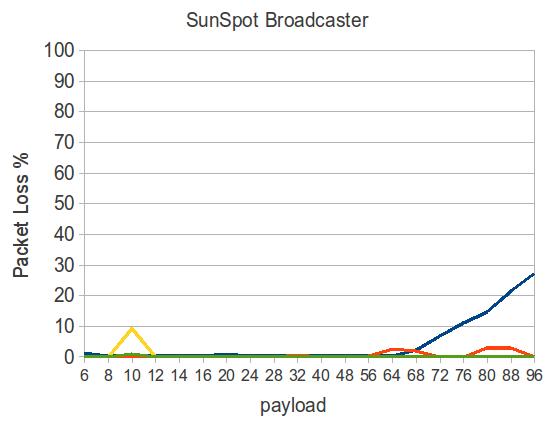}
\includegraphics[width=0.32\textwidth]{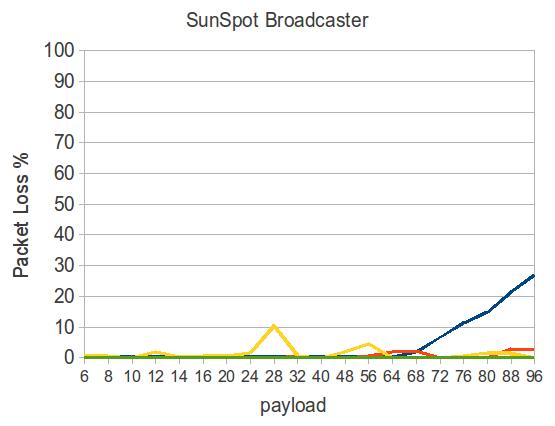}
\includegraphics[width=0.32\textwidth]{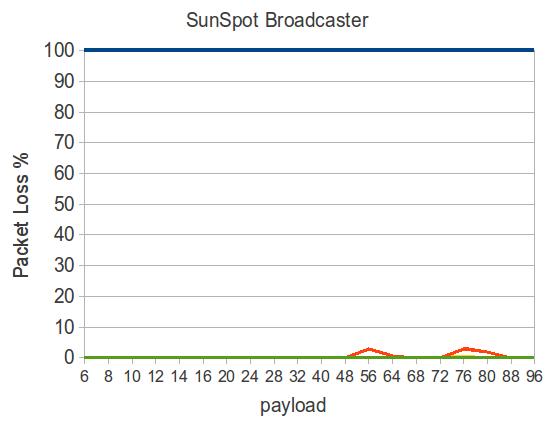}
\includegraphics[width=0.32\textwidth]{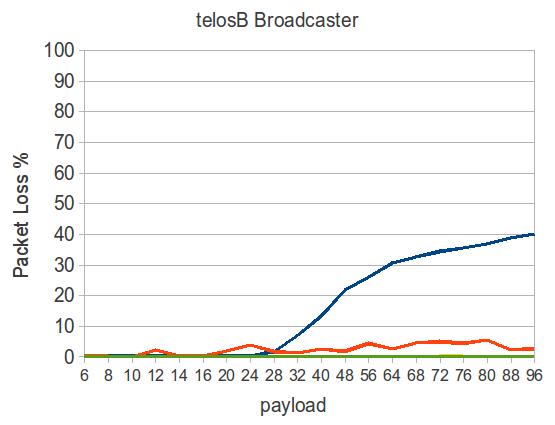}
\includegraphics[width=0.32\textwidth]{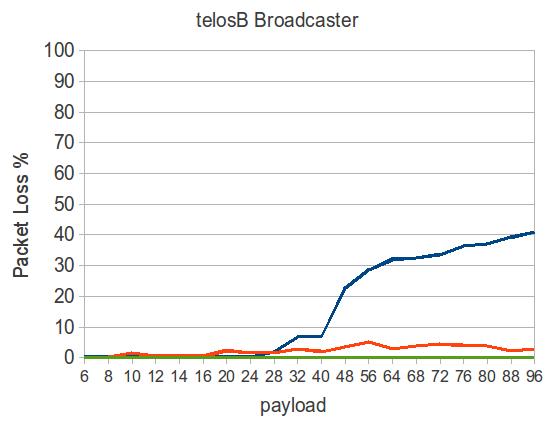}
\includegraphics[width=0.32\textwidth]{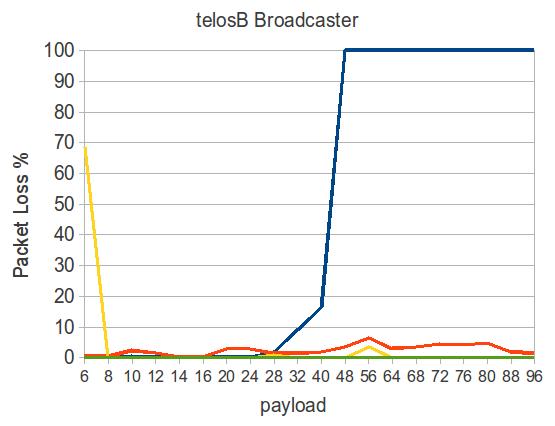}
\includegraphics[width=0.32\textwidth]{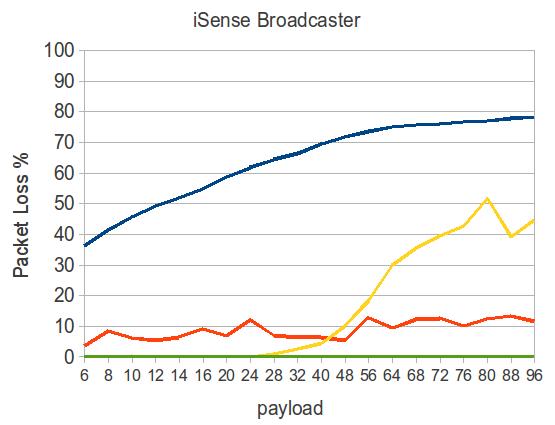}
\includegraphics[width=0.32\textwidth]{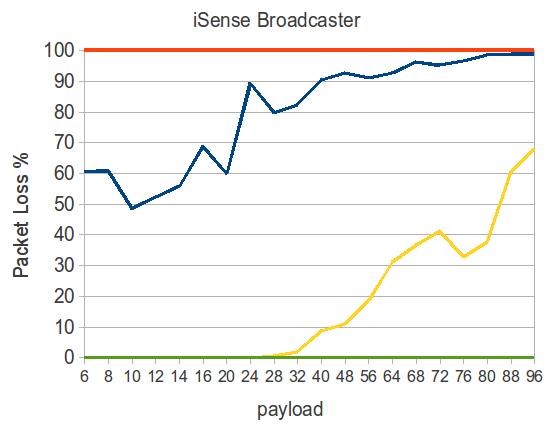}
\includegraphics[width=0.32\textwidth]{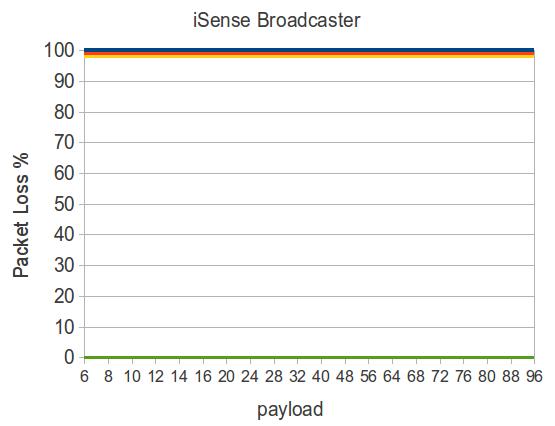}
\begin{tabular}{ m{0.1\textwidth} m{0.30\textwidth}  m{0.30\textwidth} m{0.32\textwidth}}
  & \textbf{1 meter} &\textbf{3 meters} & \textbf{8.5 meters}
\end{tabular}
\includegraphics[width=0.4\textwidth]{labels.jpg}
\caption{\label{packetloss} Packet Loss Rate}
\end{figure}

In the following figure, the Packet loss rates are illustrated, figure~\ref{packetloss}.

\newpage

Arduino has the larger packet loss rate, because of its limited receiving rate.
Moreover, although the SunSPOT has also a limited receiving rate, the packet loss is smaller. This is due to internal Buffers on the SunSPOT's JVM, temporary store the received messages, avoiding delivery failures.
 
An interesting fact is observed, when the distance between Receiving and Transmitting Nodes is 8.5 meters and the Transmitting Node is an iSense.
Beside the receiving iSense, the rest of the Receiving Devices, have packet loss rate 100\%.
iSense is the only device with 0\% packet loss, regardless of the broadcasting device, the distance and the
packet size.

\subsection{RSSI}
RSSI values depend on the modulation format. Despite the fact that, all devices use the same modulation format (i.e., offset quadrature phase-shift keying (OQPSK) ), 
the RSSI values present great deviation on identical signals; that is the same transmitted signal, simultaneously received from all Receiving Nodes.
This phenomenon is also observed on SunSPOT and TelosB, despite the fact that both nodes utilize the same CC2420 transceiver.

The RSSI.RSSI\_VAL register on CC2420~\cite{cc2420} chip, stores a digital 8 bit, signed 2's complement value.
The RSSI value is always averaged over 8 symbol periods (128  $\mu$s), as defined in 802.15.4 standard.
The RSSI.RSSI\_VAL  value can be referred to the power P using the following equations:\\
$P = RSSI\_VAL + RSSI\_OFFSET [dBm]$, where the RSSI\_OFFSET is found empirically during system development.\\
RSSI\_OFFSET is approximately -45. E.g., if reading a value of -50 from the RSSI\_VAL register, the RSSI value is approximately -5.

The problem appears in the function which returns the value of the RSSI from a received packet, on TelosB. The specific function omits the RSSI\_OFFSET and returns the raw RSSI\_VAL. Moreover, the function does not apply the 2's complement and as a result returns a wrong RSSI value. On the other hand, SunSPOT implementation follows exactly the above specifications.

In the following figure, the RSSI values are illustrated, figure~\ref{rssi}.

\begin{figure}[!ht]
\centering
\includegraphics[width=0.32\textwidth]{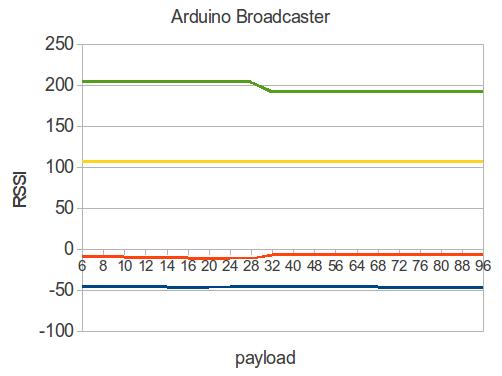}
\includegraphics[width=0.32\textwidth]{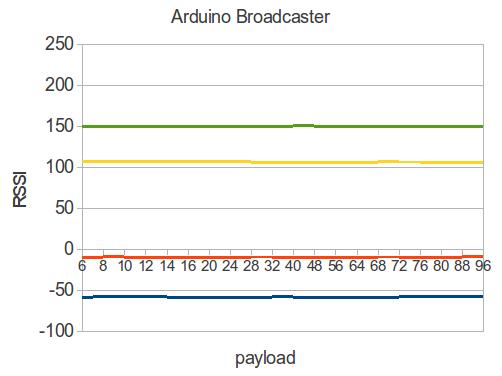}
\includegraphics[width=0.32\textwidth]{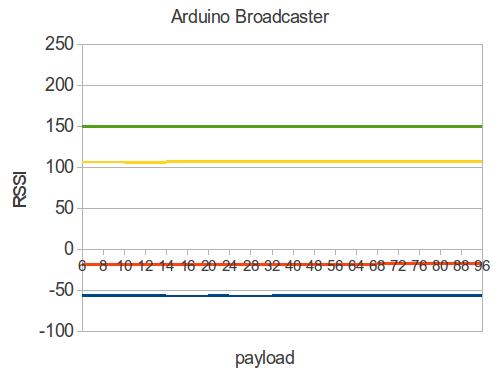}
\includegraphics[width=0.32\textwidth]{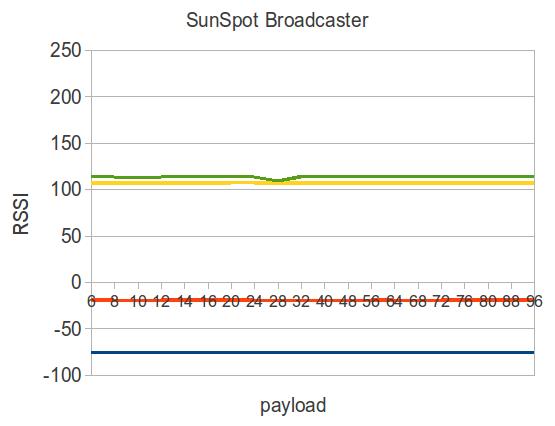}
\includegraphics[width=0.32\textwidth]{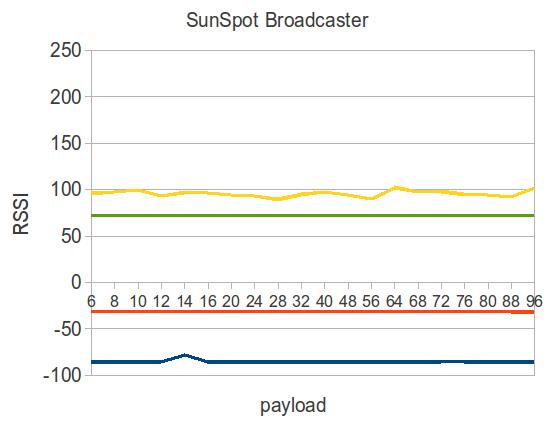}
\includegraphics[width=0.32\textwidth]{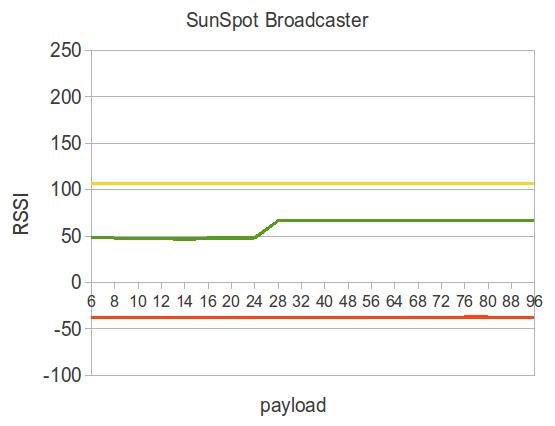}
\includegraphics[width=0.32\textwidth]{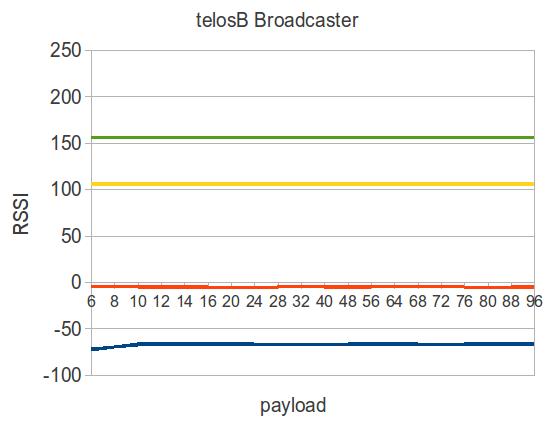}
\includegraphics[width=0.32\textwidth]{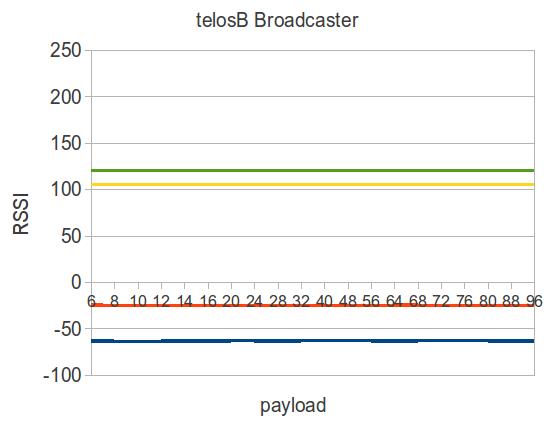}
\includegraphics[width=0.32\textwidth]{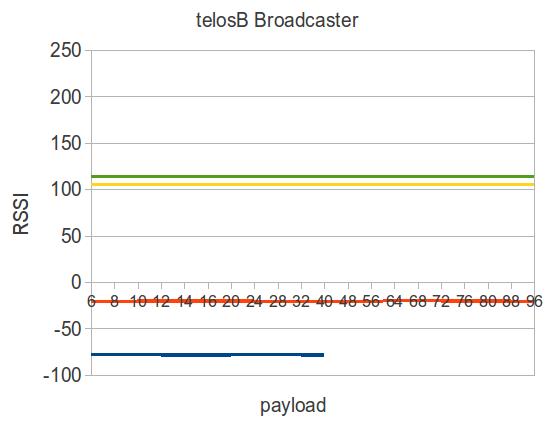}
\includegraphics[width=0.32\textwidth]{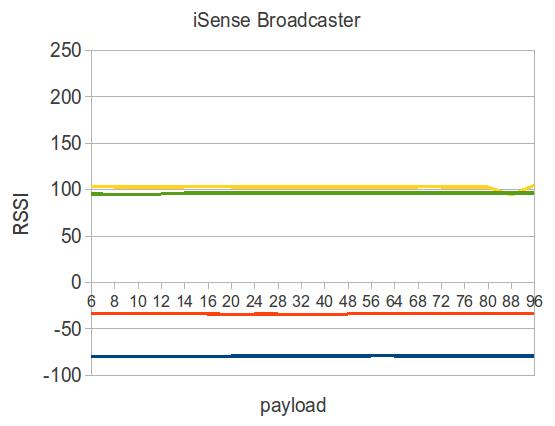}
\includegraphics[width=0.32\textwidth]{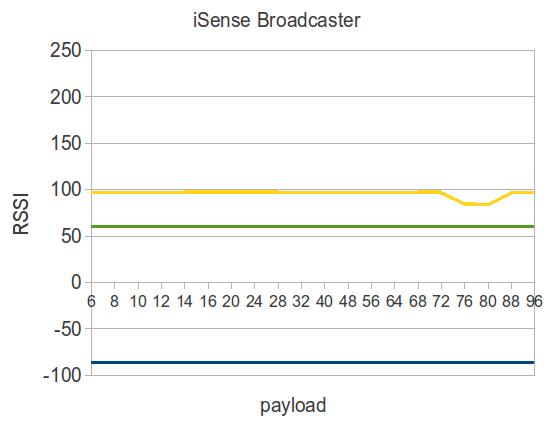}
\includegraphics[width=0.32\textwidth]{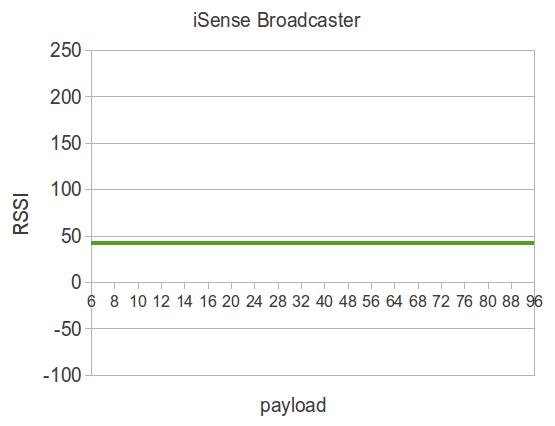}
\begin{tabular}{ m{0.1\textwidth} m{0.30\textwidth}  m{0.30\textwidth} m{0.32\textwidth}}
  & \textbf{1 meter} &\textbf{3 meters} & \textbf{8.5 meters}
\end{tabular}
\includegraphics[width=0.4\textwidth]{labels.jpg}
\caption{\label{rssi} RSSI values}
\end{figure}

\newpage

\section{Conclusions}
Studying the results presented above we conclude that the sensor node with
the
larger broadcasting and receiving rate is the iSense. iSense is the only device
with 0\% packet loss, regardless of the broadcasting device, the distance and the
packet size.

In contrast to iSense, Arduino is the only device with packet loss over 50\% in
large distances, with packet size over 50 bytes.
Moreover the broadcasting and receiving rate on Arduino, is very limited.
This might be explained by the fact that Arduino and the XBee module are
connected via the UART interface, in contrast to iSense, TelosB and SunSPOT
in which the radio module is connected to the core sensor module via the SPI
interface.

The rate of broadcasting on SunSPOT is constant. SunSPOT is running the
Squawk Java Virtual Machine, which limits the performance  of the device. The
overhead on the specific device is on the construction and the destruction of
``Datagram Java objects'' to 802.15.4 frames and vise versa and is confirmed by
the
fact that TelosB, with the same radio transceiver (CC2420), achieves much higher
transmission rates.

Regarding the RSSI values on the Receiving Nodes, there is no clear
conclusion  as to the correlation of the specific values. Each of the 4
Receiving Nodes extract, totally different RSSI values from broadcasted packets on same
distances.

Our future work concerns the examination of the correlation between RSSI and LQI values in a Heterogeneous Network. 
Also, we want to examine the performance of the XBee module when it is connected to  PC via a Serial to USB regulator. 
Moreover, it would be interesting to study the behavior of various algorithms (neighbor discovery, routing, etc.)  in such Heterogeneous Networks.
\clearpage


	\label{lastpage}
\bibliography{mksense}

\begin{thebibliography}{9}

\bibitem{arduino}
  Arduino website, http://www.arduino.cc/

\bibitem{waspmote}
  Waspomote website, http://www.libelium.com/

\bibitem{prismasense}
  Prisma Sense Quax MS-Pro website, http://www.prisma.gr/

\bibitem{scatter}
  ScatterWeb MSB-430 website, http://cst.mi.fu-berlin.de/projects/ScatterWeb/

\bibitem{isense}
  iSense website, http://www.coalesenses.com/

\bibitem{sunspot}
  SunSPOT website, http://www.sunspotworld.com/

\bibitem{telosb}
  TelosB website, http://www.xbow.com


\bibitem{xbee}
  Xbee module website, http://www.digi.com/

\bibitem{802}  
  IEEE std. 802.15.4 - 2003: Wireless Medium Access Control (MAC) and
  Physical Layer (PHY) specifications for Low Rate Wireless Personal Area
  Networks (LR-WPANs),
http://standards.ieee.org/getieee802/download/802.15.4-2003.pdf

\bibitem{tinyos}    
The TinyOS website, http://www.tinyos.net

\bibitem{tinyspot}  
Daniel van den Akker, Kurt Smolderen, Peter De Cleyn, Bart Braem, and Chris Blondia: TinySPOTComm: Facilitating communication
over IEEE 802.15.4 between Sun SPOTs and TinyOS-based motes.

\bibitem{wiring}  
 Wiring website, http://wiring.org.co/

\bibitem{processing}  
 Processing website, http://processing.org/

\bibitem{cc2420}  
CC2420 transceiver datasheet, http://focus.ti.com/lit/ds/symlink/cc2420.pdf

\end{thebibliography}

\end{document}